\documentclass{elsart}
\usepackage{amssymb}
\usepackage{graphicx}
\begin{document}
\begin{frontmatter}
\title{Thermodynamic magnetization of a strongly correlated two-dimensional electron system}
\author{S.~V. Kravchenko,\corauthref{cor1}}
\corauth[cor1]{Corresponding author}
\ead{s.kravchenko@neu.edu}
\author{A.~A. Shashkin,\thanksref{a}}
\thanks[a]{Permanent address: Institute of Solid State Physics, Chernogolovka, Moscow District 142432, Russia}
\author{S. Anissimova,}
\author{A. Venkatesan,\thanksref{b}}
\thanks[b]{Present address: Department of Physics and Astronomy, University of British Columbia, Vancouver, BC V6T~1Z1, Canada}
\author{M.~R. Sakr\thanksref{c}}
\thanks[c]{Present address: Department of Physics and Astronomy, UCLA, Los Angeles, CA 90095, USA}
\address{Physics Department, Northeastern University, Boston, Massachusetts 02115, USA}
\author{V.~T. Dolgopolov}
\address{Institute of Solid State Physics, Chernogolovka, Moscow District 142432, Russia}
\author{T.~M. Klapwijk}
\address{Kavli Institute of Nanoscience, Delft University of Technology, 2628 CJ Delft, The Netherlands}

\begin{abstract}
We measure thermodynamic magnetization of a low-disordered, strongly
correlated two-dimensional electron system in silicon.  Pauli spin
susceptibility is observed to grow critically at low electron
densities --- behavior that is characteristic of the existence of a
phase transition.  A new, parameter-free method is used to directly
determine the spectrum characteristics (Land\'e $g$-factor and the
cyclotron mass) when the Fermi level lies outside the spectral gaps
and the inter-level interactions between quasiparticles are avoided.
It turns out that, unlike in the Stoner scenario, the critical growth
of the spin susceptibility originates from the dramatic enhancement
of the effective mass, while the enhancement of the $g$-factor is
weak and practically independent of the electron density.
\end{abstract}

\end{frontmatter}

\section{Introduction}\vspace{-5mm}
\enlargethispage{5mm}
Presently, theoretical description of interacting electron systems is
restricted to two limiting cases: (i)~weak electron-electron
interactions (small ratio of the Coulomb and Fermi energies
$r_s=E_C/E_F\ll1$, high electron densities) and (ii)~very strong
electron-electron interactions ($r_s\gg1$, very low electron
densities). In the first case, conventional Fermi-liquid behavior
\cite{landau} is established, while in the second case, formation of
the Wigner crystal is expected \cite{wigner} (for recent
developments, see Ref.~\cite{tanatar}). Numerous experiments
performed in both three- (3D) and two-dimensional (2D) electron
systems at intermediate interaction strengths ($1\lesssim
r_s\lesssim5$) have not demonstrated any significant change in
properties compared to the weakly-interacting regime (see, {\it
e.g}., Refs.~\cite{3d,2d}). It was not until recently that
qualitative deviations from the weakly-interacting Fermi liquid
behavior (in particular, the drastic increase of the effective
electron mass with decreasing electron density) have been found in
strongly correlated 2D electron systems ($r_s\gtrsim10$)
\cite{review1}. However, these findings have been based solely on the
studies of a kinetic parameter (conductivity), which, in general, is
not a characteristic of a state of matter.

The 2D electron system in silicon turns out to be a very convenient
object for studies of the strongly correlated regime due to the large
interaction strengths ($r_s>10$ can be easily reached) and high
homogeneity of the samples estimated (from the width of the
magnetocapacitance minima in perpendicular magnetic fields) at about
$4\times10^9$~cm$^{-2}$ \cite{shashkin01}. Here we report
measurements of the thermodynamic magnetization and density of states
in such a low-disordered, strongly correlated 2D electron system in
silicon. We concentrate on the metallic regime where conductivity
$\sigma\gg e^2/h$. We have found that in this system subjected to
parallel magnetic fields, the spin susceptibility of band electrons
(Pauli spin susceptibility) becomes enhanced by almost an order of
magnitude at low electron densities, growing critically near a
certain critical density $n_\chi\approx8\times10^{10}$~cm$^{-2}$:
behavior that is characteristic in the close vicinity of a phase
transition. The density $n_\chi$ is coincident within the
experimental uncertainty with the critical density $n_c$ for the
zero-field metal-insulator transition (MIT) in our samples. We have
also found by measurements in perpendicular and tilted magnetic
fields that the $g$-factor is weakly enhanced and practically
independent of the electron density down to the lowest densities
reached, while the cyclotron mass becomes strongly enhanced at low
$n_s$. Thus, unlike in the Stoner scenario, it is the effective mass,
rather than the g-factor, that is responsible for the dramatically
enhanced spin susceptibility at low electron densities.

\section{Experimental setup and samples}\vspace{-5mm}

\begin{figure}[b]
\hspace{5.9mm}\scalebox{.54}{\includegraphics[angle=90]{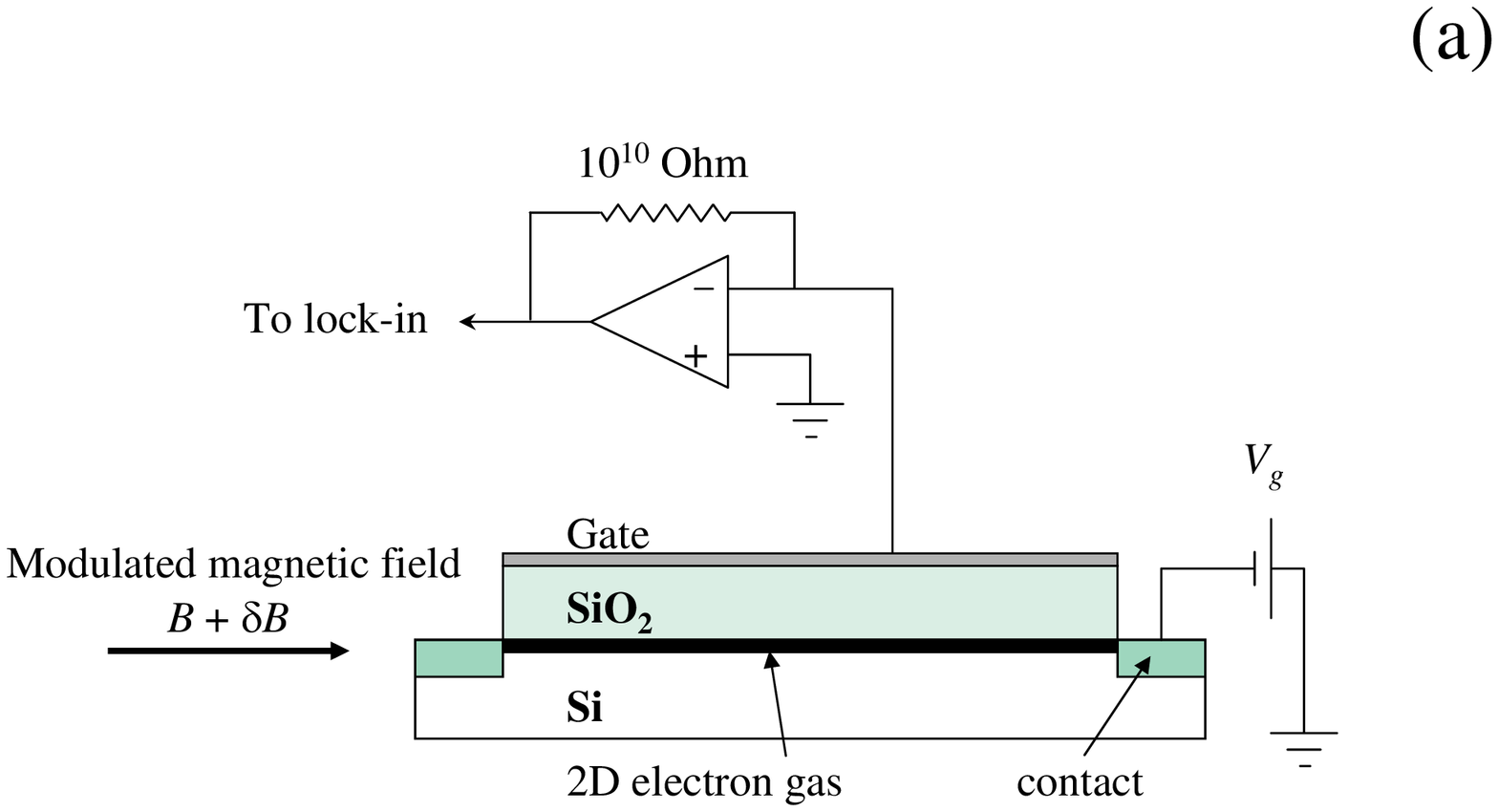}}\vspace{7mm}
    \centering
    \hspace{10mm}\scalebox{.5}{\includegraphics[angle=90]{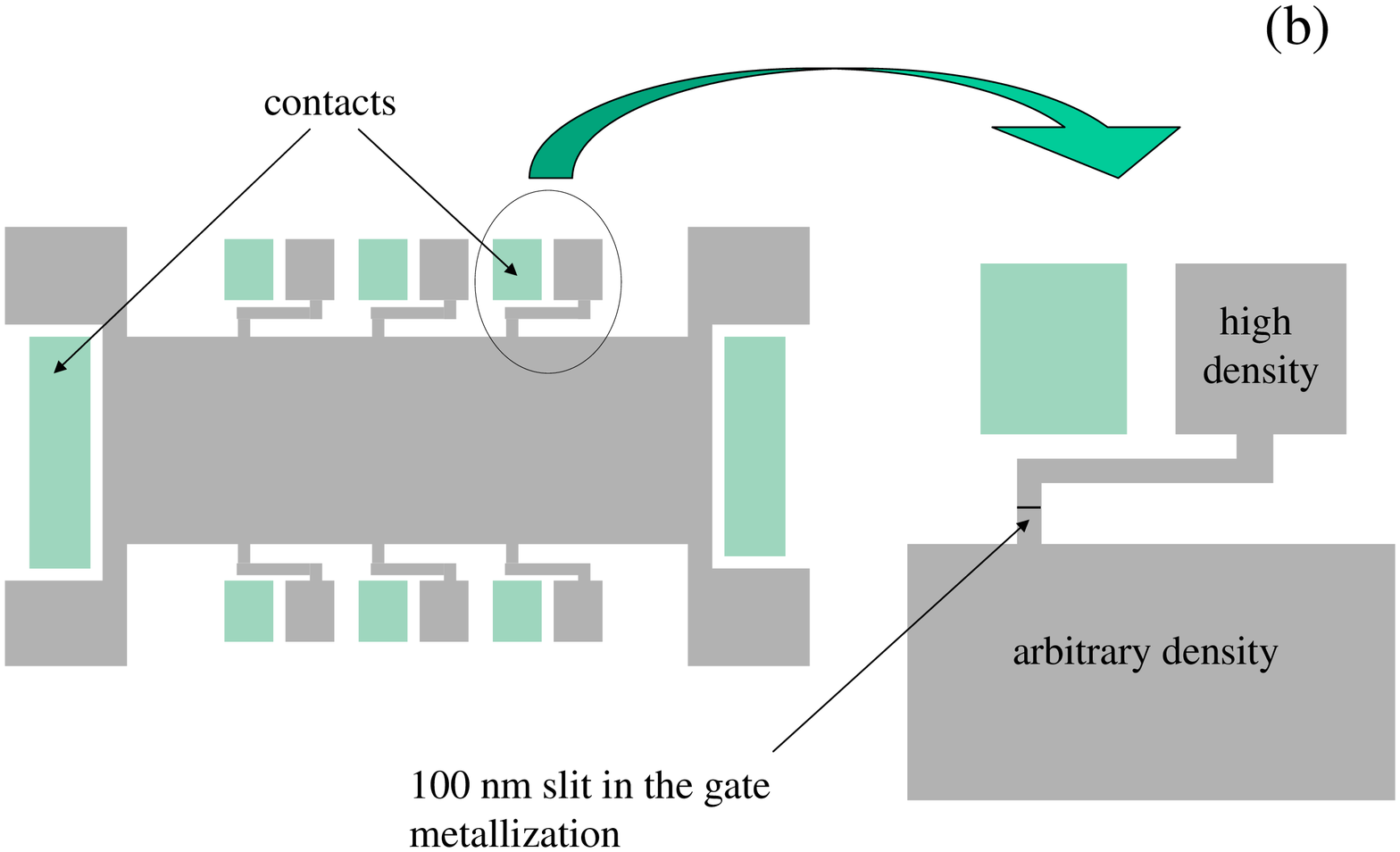}}
\caption{\label{fig1} (a)~Experimental set-up for measurements of
thermodynamic magnetization. (b)~Top view of a MOSFET with slit in
the gate metalization indicated.}
\end{figure}

Magnetization is one of the least studied properties of 2D electron
systems: signals associated with it are weak making measuring them a
challenging experiment. Few experimental observations of the
de~Haas–-van~Alphen effect in 2D electron systems were made using
SQUID magnetometers \cite{stormer83}, pick up coils lithographed
above the gate \cite{fang83}, or torque magnetometers
\cite{eisenstein85}. Here we use a novel method \cite{prus03} to
measure the magnetization that entails modulating the magnetic field
with an auxiliary coil and measuring the AC current induced between
the gate and the 2D electron system. The experimental set-up is shown
in Fig.~\ref{fig1}~(a). Magnetic field $B$ was modulated with a small
ac field $\delta B$ in the range 0.005 -- 0.03~T at frequencies
between $f=0.05$ and $0.45$~Hz.  The in-phase and out-of-phase
components of the current between the gate and the 2D electron system
were measured with high precision ($\sim10^{-16}$~A) using a
current-voltage converter and a lock-in amplifier. The real
(in-phase) component of the current depends on the dissipative
conductivity $\sigma$ of the sample. We are, however, interested in
the imaginary (out-of-phase) current component, which, under
condition that $2\pi fC\ll\sigma$, is equal to
\begin{equation}
\mbox{Im }i=\frac{2\pi fC}{e}\,\frac{d\mu}{dB}\,\delta B ,
\end{equation}
where $C$ is the capacitance of the sample measured in the same
experiment and $\mu$ is the chemical potential. Use of ultra-low
frequencies ensures that $2\pi fC\ll\sigma$ and the out-of-phase
current component is not contaminated by lateral transport effects.
By applying the Maxwell relation $dM/dn_s=-d\mu/dB$, one can then
obtain the magnetization $M$ from the measured $i$. For measurements
of the thermodynamic density of states, a similar circuit was used
with a distinction that the gate voltage was modulated and thus the
imaginary current component was proportional to the capacitance.
Thermodynamic density of states $dn_s/d\mu$ is related to
magnetocapacitance via
\begin{equation}
\frac{1}{C}=\frac{1}{C_0}+\frac{1}{Ae^2(dn_s/d\mu)},
\end{equation}
where $C_0$ is the geometric capacitance and $A$ is the sample area.

Measurements were made in an Oxford dilution refrigerator on
low-disordered (100)-silicon metal-oxide-semiconductor field-effect
transistors (MOSFETs).  These samples are remarkable by the absence
of a band tail of localized electrons down to electron densities
$n_s\approx1\times10^{11}$~cm$^{-2}$~\cite{review1}, which
allows one to study properties of a {\em clean} 2D electron system
without admixture of local moments \cite{mott,moments,jp}.  The level
of disorder is determined by the quality of the Si-SiO$_2$ interface.
The silicon oxide layer in our samples was grown repeatedly during
several stages of the fabrication to decrease number of charged
impurities and obtain high quality interface; as a result, in samples
with oxide thickness of 149~nm, we have achieved peak mobilities of
3~m$^2$/Vs at $T=100$~mK. The second advantage of these samples is a
very low contact resistance (in ``conventional'' silicon samples,
high contact resistance becomes the main experimental obstacle in the
low density/low temperature limit). To minimize contact resistance,
thin slits in the gate metalization have been introduced (see
Fig.~\ref{fig1}~(b)), which allows for maintaining high electron
density near the contacts regardless of its value in the main part of
the sample.

\begin{figure}
\centering
\scalebox{0.5}{\includegraphics[clip]{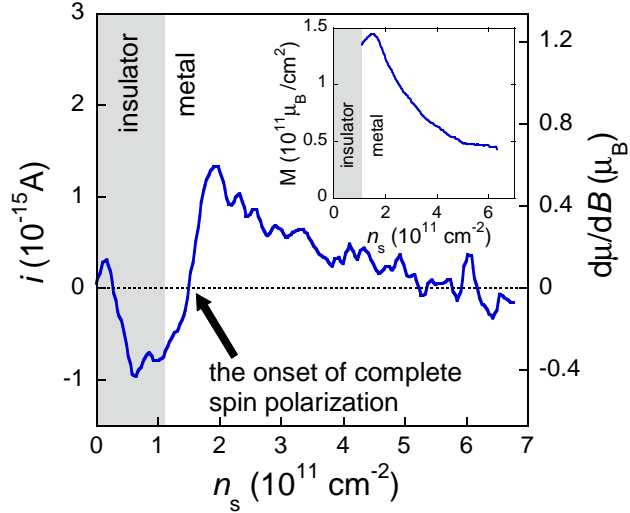}}
\caption{\label{fig2} Imaginary current component in the
magnetization experiment as a function of the electron density in a
magnetic field of 5~T and $T=0.4$~K. Grey area depicts the insulating
phase. Magnetization versus $n_s$ is displayed in the inset. Note
that the maximum $M$ is coincident within the experimental
uncertainty with $\mu_Bn_s$.}
\end{figure}

\section{Experimental results and discussion}
\subsection{Pauli spin susceptibility}

We start with measurements of Pauli spin susceptibility.  A typical
experimental trace of $i(n_s)$ in a parallel magnetic field of 5~T is
displayed in Fig.~\ref{fig2}. The inset shows magnetization $M(n_s)$
in the metallic phase obtained by integrating $dM/dn_s=-d\mu/dB$ with
the integration constant $M(\infty)=B\chi_0$, where $\chi_0$ is the
Pauli spin susceptibility of non-interacting electrons. A nearly
anti-symmetric jump of $i(n_s)$ about zero on the $y$-axis (marked by
the black arrow) separates the high- and low-density regions in which
the signal is positive and negative ($M(n_s)$ is decreasing and
increasing), respectively. Such a behavior is expected based on
simple considerations. At low densities, all electrons are
spin-polarized in a magnetic field, so for the simple case of
non-interacting 2D electrons one expects $d\mu/dB=-\mu_B$ (at
$n_s\rightarrow0$, deep in the insulating regime, the capacitance of
the system vanishes and, therefore, the measured current approaches
zero). At higher densities, when the electrons start to fill the
upper spin subband, $M(n_s)$ starts to decrease, and $d\mu/dB$ is
determined by the renormalized Pauli spin susceptibility $\chi$ and
is expected to decrease with $n_s$ due to reduction in the strength
of electron-electron interactions. Finally, in the high-density
limit, the spin susceptibility approaches its ``non-interacting''
value $\chi_0$, and $d\mu/dB$ should approach zero. The onset of
complete spin polarization --- the electron density $n_p$ at which
the electrons start to fill the upper spin subband --- is given by
the condition $d\mu/dB=0$ ($M(n_s)$ reaches a maximum), as indicated
by the black arrow in the figure. It is important that over the range
of magnetic fields used in the experiment (1.5--7~tesla), the maximum
$M$ coincides within the experimental uncertainty with $\mu_Bn_s$
thus confirming that all the electrons are indeed spin-polarized
below $n_p$. Note however that the absolute value of $d\mu/dB$ at
$n_s\lesssim n_c$ is reduced in the experiment. We attribute this to
smearing of the minimum in $i(n_s)$ caused by possible influence of
the residual disorder in the electron system, which is crucial in and
just above the insulating phase, in contrast to the clean metallic
regime we focus on here. Another reason for the reduction in
$d\mu/dB$ is the electron-electron interactions (due to, {\it e.g.},
the enhanced effective mass).

\begin{figure}\centering
\scalebox{0.6}{\includegraphics[clip]{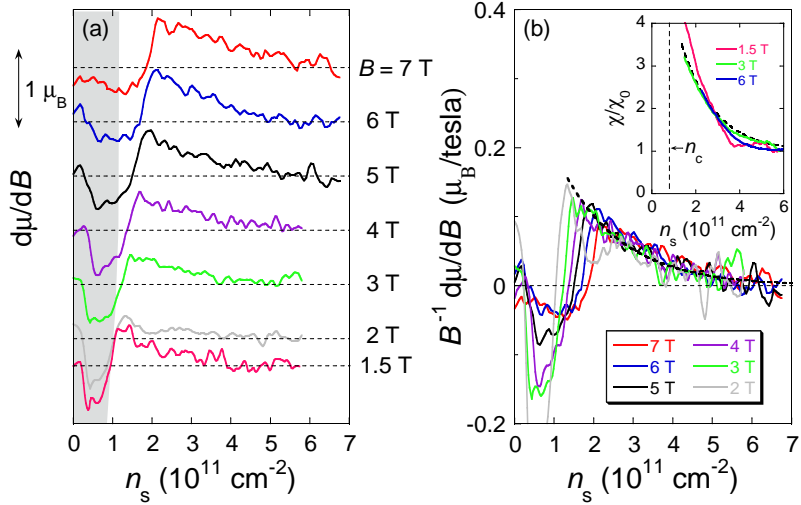}}
\caption{\label{fig3} (a)~The experimental $d\mu/dB$ as a function of
electron density in different magnetic fields and $T=0.4$~K. The
curves are vertically shifted for clarity. Grey area depicts the
insulating phase. Note that the onset of full spin polarization in
our experiment always takes place in the metallic regime. (b)~Scaling
of the $d\mu/dB$ curves, normalized by magnetic field magnitude, at
high electron densities. The dashed line represents the ``master
curve''. Spin susceptibility obtained by integrating the master curve
(dashed line) and raw data at $B=$~1.5, 3, and 6~T is displayed in
the inset.}
\end{figure}

In Fig.~\ref{fig3}(a), we show a set of curves for the experimental
$d\mu/dB$ versus electron density in different magnetic fields.
Experimental results in the range of magnetic fields studied do not
depend, within the experimental noise, on temperature below 0.6~K
(down to 0.15~K which was the lowest temperature achieved in this
experiment). The onset of full spin polarization shifts to higher
electron densities with increasing magnetic field. Grey area depicts
the insulating phase, which expands somewhat with $B$ (for more on
this, see Ref.~\cite{MIT}). Note that the range of magnetic fields
used in our experiment is restricted from below by the condition that
$d\mu/dB$ crosses zero in the metallic regime. In Fig.~\ref{fig3}(b),
we show how these curves, normalized by magnetic field, collapse in
the partially-polarized regime onto a single ``master curve''. The
existence of such scaling verifies proportionality of the
magnetization to $B$, confirming that we deal with Pauli spin
susceptibility of band electrons, and establishes a common zero level
for the experimental traces. Integration of the master curve over
$n_s$ yields the spin susceptibility $\chi=M/B$, as shown in the
inset to Fig.~\ref{fig3}(b). Also shown is the spin susceptibility
obtained by integration of raw curves at $B=$~1.5, 3, and 6 tesla,
which, within the experimental error, yield the same dependence.

This method of measuring the spin susceptibility, being the most
direct, suffers, however, from possible influence of the unknown
diamagnetic contribution to the measured $d\mu/dB$, which arises from
the finite width of the 2D electron layer \cite{rem}. To verify that
this influence is negligible in our samples, we employ another two
independent methods to determine $\chi$. The second method is based
on marking the electron density $n_p$ at which $d\mu/dB=0$ and which
corresponds to the onset of complete spin polarization, as mentioned
above. The so-determined polarization density $n_p(B)$ can be easily
converted into $\chi(n_s)$ via $\chi=\mu_Bn_p/B$. Note that in
contrast to the value of $d\mu/dB$, the polarization density $n_p$ is
practically not affected by possible influence of the diamagnetic
shift.

\begin{figure}\centering
\scalebox{0.8}{\includegraphics[clip]{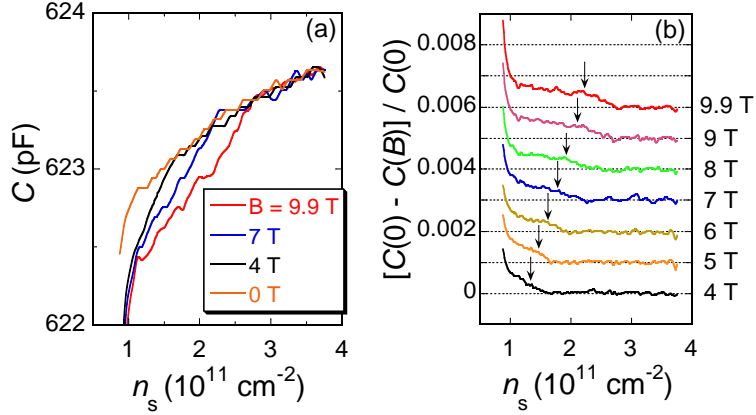}}
\caption{\label{fig4} (a)~Magnetocapacitance versus electron density
for different magnetic fields. (b)~Deviation of the $C(n_s)$
dependences for different magnetic fields from the $B=0$ reference
curve. The traces are vertically shifted for clarity. The onset of
full spin polarization is indicated by arrows.}
\end{figure}

The third method for measuring $n_p$ and $\chi$, insensitive to the
diamagnetic shift, relies on analyzing the magnetocapacitance, $C$.
Experimental traces $C(n_s)$ are shown in Fig.~\ref{fig4}(a) for
different magnetic fields. As the magnetic field is increased, a
step-like feature emerges on the $C(n_s)$ curves and shifts to higher
electron densities. This feature corresponds to the thermodynamic
density of states abruptly changing when the electrons' spins become
completely polarized. To see the step-like feature more clearly, in
Fig.~\ref{fig4}(b) we subtract the $C(n_s)$ curves for different
magnetic fields from the reference $B=0$ curve. The fact that the
jumps in $C$ (as well as in $d\mu/dB$) are washed out much stronger
than it can be expected from possible inhomogeneities in the electron
density distribution (about $4\times10^9$~cm$^{-2}$
\cite{shashkin01}) points to the importance of electron-electron
interactions. Since the effects of interactions are different in the
fully- and partially-polarized regimes, it is natural to mark the
onset of full spin polarization at the beginning of the
interaction-broadened jump, as indicated by arrows in the figure. In
case the residual disorder does contribute to the jump broadening, we
extend error bars to the middle of the jump, which yields an upper
boundary for the onset of full spin polarization.

\begin{figure}\centering
\scalebox{0.5}{\includegraphics[clip]{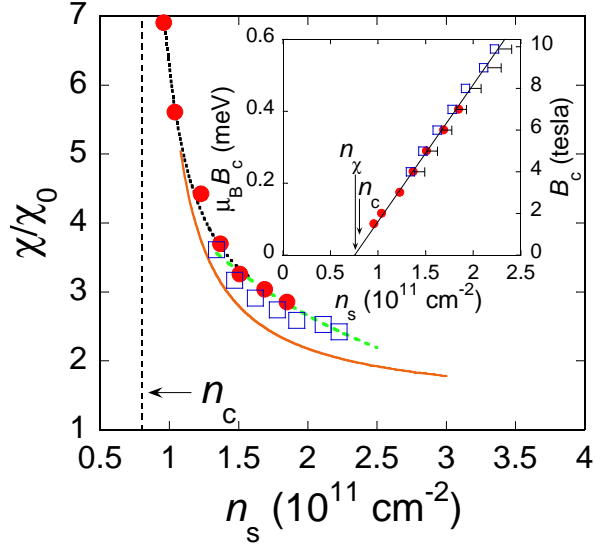}}
\caption{\label{fig5} Dependence of the Pauli spin susceptibility on
electron density obtained by all three methods described in text:
integral of the master curve (dashed line), $d\mu/dB=0$ (circles),
and magnetocapacitance (squares). The dotted line is a guide to the
eye. Also shown by a solid line is the transport data of
Ref.~\cite{shashkin01}. Inset: polarization field as a function of
the electron density determined from the magnetization (circles) and
magnetocapacitance (squares) data. The symbol size for the
magnetization data reflects the experimental uncertainty, and the
error bars for the magnetocapacitance data extend to the middle of
the jump in $C$. The data for $B_c$ are described by a linear fit
which extrapolates to a density $n_\chi$ close to the critical
density $n_c$ for the $B=0$ MIT.}
\end{figure}

In Fig.~\ref{fig5}, we show the summary of the results for the Pauli
spin susceptibility as a function of $n_s$, obtained using all three
methods described above. The excellent agreement between the results
obtained by all of the methods establishes that a possible influence
of the diamagnetic shift is negligible \cite{remark} and, therefore,
the validity of the data including those at the lowest electron
densities is justified. There is also good agreement between these
results and the data obtained by the transport experiments of
Ref.~\cite{shashkin01}. This adds credibility to the transport data
and confirms that full spin polarization occurs at the field $B_c$;
however, we note again that evidence for the phase transition can
only be obtained from thermodynamic measurements. The magnetization
data extend to lower densities than the transport data, and larger
values of $\chi$ are reached, exceeding the ``non-interacting'' value
$\chi_0$ by almost an order of magnitude. The Pauli spin
susceptibility behaves critically close to the critical density $n_c$
for the $B=0$ metal-insulator transition \cite{rem1}: $\chi\propto
n_s/(n_s-n_\chi)$. This is in favor of the occurrence of a
spontaneous spin polarization (either Wigner crystal \cite{rem2} or
ferromagnetic liquid) at low $n_s$, although in currently available
samples, the formation of the band tail of localized electrons at
$n_s\lesssim n_c$ conceals the origin of the low-density phase. In
other words, so far, one can only reach an incipient transition to a
new phase.

The dependence $B_c(n_s)$, determined from the magnetization and
magnetocapacitance data, is represented in the inset to
Fig.~\ref{fig5}. The two data sets coincide and are described well by
a common linear fit which extrapolates to a density $n_\chi$ close to
$n_c$. We emphasize that in the low-field limit ($B<1.5$~tesla), the
jump in $d\mu/dB$ shifts to the insulating regime, which does not
allow us to approach closer vicinity of $n_\chi$: based on the data
obtained in the regime of strong localization, one would not be able
to make conclusions concerning properties of a clean metallic
electron system which we are interested in here. Clearly, the fact
that the linear $B_c(n_s)$ dependence persists down to the lowest
electron densities achieved in the experiment confirms that we always
deal with the clean metallic regime.

In the end of this subsection, we would like to clarify the principal
difference between our results and those of Ref.~\cite{prus03}. In
the sample used by Prus {\it et al}., the critical density $n_c$ for
the $B=0$ MIT was considerably higher than in our samples caused by
high level of disorder, and the band tail of localized electrons was
present at all electron densities \cite{prus03}. As a result, the
crucial region of low electron densities, in which the critical
behavior of the Pauli spin susceptibility occurs, falls within the
insulating regime where the physics of local moments dominates
\cite{mott,moments,jp}. Indeed, Prus {\it et al}.\ have found
sub-linear $M(B)$ dependence characteristic of local moments, and the
extracted spin susceptibility in their sample has a Curie temperature
dependence \cite{jp}. This is the case even at high electron
densities, where metallic behavior might be expected instead. Such
effects are absent in our samples: the spin susceptibility (in the
partially-polarized system) is independent of the magnetic field and
temperature, confirming that we deal with Pauli spin susceptibility
of band electrons.

\subsection{Magnetization in perpendicular magnetic fields: $g$-factor and effective mass}

\begin{figure}\centering
\scalebox{0.5}{\includegraphics[clip]{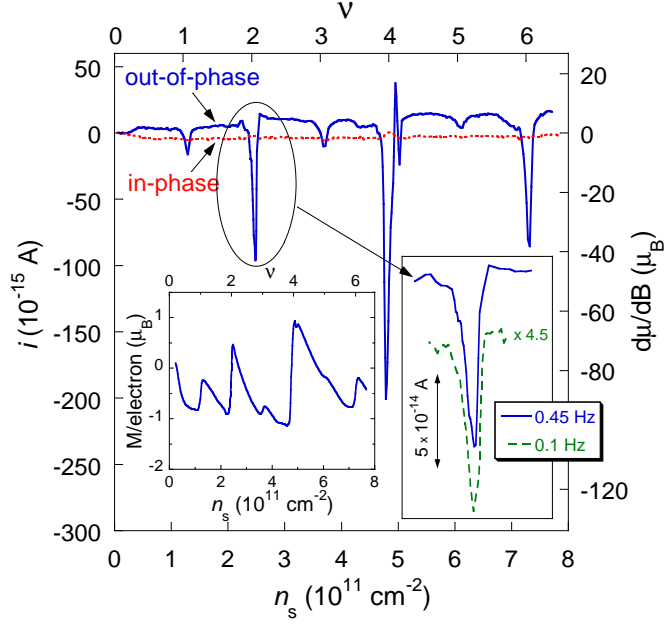}}
\caption{\label{fig6} Out-of-phase (solid line) and in-phase (dotted
line) current components as a function of the electron density in a
perpendicular magnetic field of 5~T and $T=0.8$~K. $\delta B=0.022$~T
and $f=0.45$~Hz. The value $d\mu/dB$ is indicated in units of the
Bohr magneton $\mu_B$. In the right-hand inset, we demonstrate
proportionality of $\mbox{Im }i$ to frequency: the solid and dashed
lines (vertically shifted for clarity) correspond to $0.45$ and
$0.1$~Hz, respectively; the $y$-component of the latter is multiplied
by 4.5. The left-hand inset illustrates magnetization per electron.}
\end{figure}

Typical experimental traces of the gate current in a perpendicular
magnetic field of 5~T are displayed in Fig.~\ref{fig6}. Sharp dips in
the out-of-phase component, seen at integer filling factors
$\nu\equiv n_shc/eB_\perp$, reflect gaps in the density of states:
dips at odd filling factors correspond to the valley splitting, the
ones at $\nu=2$ and 6 are due to the spin splitting, and the dip at
$\nu=4$ is due to the cyclotron splitting. However, there are no
corresponding features in the in-phase current component, which
ensures that we reach the low-frequency limit and the measured
$\partial\mu/\partial B$ is not distorted by lateral transport
effects. This is further confirmed by the fact that the out-of-phase
current is proportional to the excitation frequency as displayed in
the right-hand inset to Fig.~\ref{fig6}. Magnetization per electron
can be extracted by integrating the measured out-of-phase signal with
respect to $n_s$, as shown in the left-hand inset to Fig.~\ref{fig6}
for illustration. The magnetization exhibits the expected sawtooth
oscillations, with sharp jumps at integer filling factors (note that
the height of the jumps yields values that are smaller than the level
splitting by the level width).

If the disorder and interactions are disregarded, in quantizing
magnetic fields (except at integer filling factors) the derivative
$\partial\mu/\partial B=-\partial M/\partial n_s$ is equal to
\begin{equation}
\frac{\partial\mu}{\partial B}=\mu_B\left[\left(\frac{1}{2}+N\right)\frac{2m_e}{m_b}\pm\frac{1}{2}g_0\right],
\end{equation}
where $\mu_B$ is the Bohr magneton, $N$ is the Landau level number,
$m_e$ and $m_b=0.19\,m_e$ are the free electron mass and band mass,
respectively, and $g_0=2$ is the $g$-factor in bulk silicon. Disorder
smears out the dependences which otherwise would consist of a series
of delta-functions. Interactions modify this picture in two ways:
(i)~by renormalizing the values of the cyclotron mass and $g$-factor
and (ii)~by providing a negative contribution of order
$-(e^2/\varepsilon l_B)\{\nu\}^{1/2}$ to the chemical potential
\cite{macdonald86,efros88} (here $\varepsilon$ is the dielectric
constant, $l_B$ is the magnetic length, and $\{\nu\}$ is the
deviation of the filling factor from the nearest integer). The latter
effect, which is caused by the intra-level interactions between
quasiparticles, leads to the so-called negative thermodynamic
compressibility near integer filling factors predicted by Efros
\cite{efros88} and experimentally observed in
Refs.~\cite{kravchenko89,eisenstein92}.

\begin{figure}\centering
\scalebox{0.7}{\includegraphics[clip]{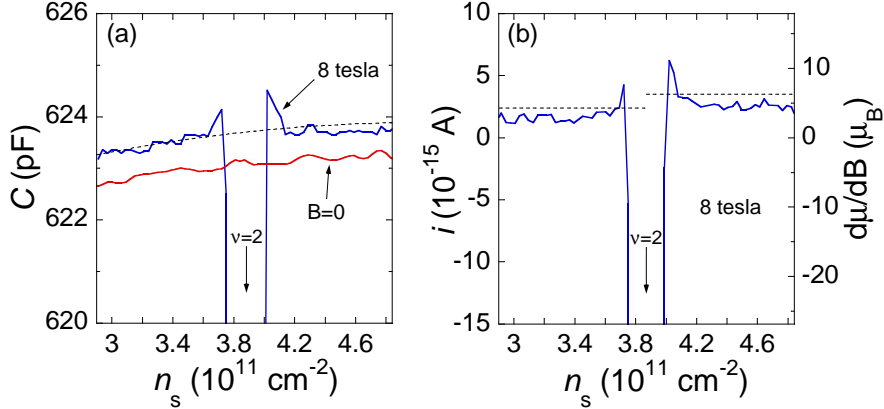}}
\caption{\label{fig7} (a)~Capacitance in $B=8$~tesla and in $B=0$ as
indicated. The (noise averaged) geometric capacitance is depicted by
a dashed line. (b)~$\mbox{Im }i\propto d\mu/dB$ in a perpendicular
magnetic field of 8~tesla. The maximum values possible in a
non-interacting system (see text) are depicted by dashed lines.}
\end{figure}

In Fig.~\ref{fig7}, we compare capacitance $C$ with
$\partial\mu/\partial B$, measured at the same magnetic field value
and plotted versus $n_s$ around the filling factor $\nu=2$. The
geometric capacitance \cite{rem3} (the inverse of the first term in
Eq.~(2)), depicted by dashed line in Fig.~\ref{fig7}(a), slightly
increases with $n_s$ since the thickness of the 2D electron layer ---
and, therefore, the average distance between the 2D layer and the
gate --- decreases with the gate voltage. The second term in Eq.~(2)
is responsible for the dip centered at
$n_s=3.87\times10^{11}$~cm$^{-2}$, corresponding to $\nu=2$, and
sharp maxima on both sides of it. Note that at these maxima, the
capacitance exceeds $C_0$, which corresponds to the negative
thermodynamic compressibility discussed above. Farther from integer
filling factors, the intra-level interaction corrections become weak,
being proportional to $\{\nu\}^{-1/2}$, and the measured capacitance
approaches $C_0$ (as long as the broadening of Landau levels is
negligible, {\it i.e.}, $dn_s/d\mu\gg\left.dn_s/d\mu\right|_{B=0}$).

Similar maxima on both sides of $\nu=2$ are seen in the magnetization
data shown in Fig.~\ref{fig7}~(b). At the maxima, the derivative
$\partial\mu/\partial B$ exceeds maximum values possible in a
non-interacting 2D electron gas, which are determined by Eq.~(3) and
are depicted in the figure by dashed lines. The possibility that
$\partial\mu/\partial B$ might exceed its maximum non-interacting
values due to intra-level Coulomb interactions between quasiparticles
was predicted by MacDonald {\it et al}.\ \cite{macdonald86}; in fact,
this is how negative compressibility \cite{efros88} manifests itself
in magnetization measurements. Sharp spike just above $\nu=4$ and
maxima on both sides of $\nu=2$ in the dependence shown in
Fig.~\ref{fig6} are of the same nature.

\begin{figure}\centering
\scalebox{0.7}{\includegraphics[clip]{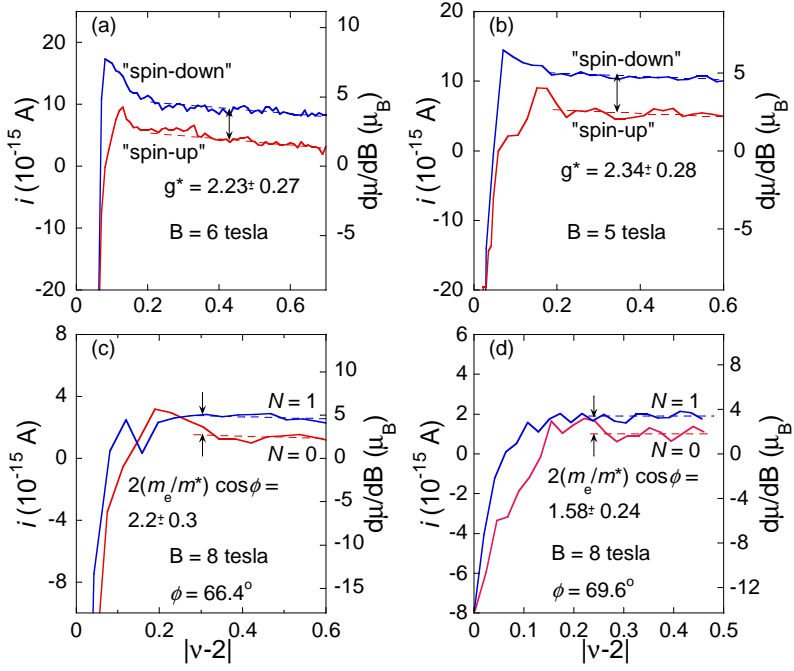}}
\caption{\label{fig8} Illustration of how the effective $g$-factor
(a, b) and the cyclotron mass (c, d) have been measured. The
imaginary current component is plotted as a function of the deviation
of the filling factor from $\nu=2$. In perpendicular magnetic fields,
the difference between $\partial\mu/\partial B$ for spin-down
($\downarrow$) and spin-up ($\uparrow$) electrons yields $g^*$ in
units of the Bohr magneton. In tilted magnetic fields, the difference
between $\partial\mu/\partial B$ for electrons with $N=1$ and $N=0$
is equal to $2\mu_B\,(m_e/m^*)\cos\,\phi$. The dashed lines show
noise-averaged values. $\delta B=0.022$~T (a, b) and 0.0055~T (c,
d).}
\end{figure}

It is straightforward to obtain the effective $g$-factor from the
data for $\partial\mu/\partial B$. In accordance with Eq.~(3), it is
equal (in units of the Bohr magneton) to the difference between
$\partial\mu/\partial B$ for spin-down ($\downarrow$) and spin-up
($\uparrow$) electrons belonging to the same Landau level:
$\mu_B\,g^*= (\partial\mu/\partial
B)_\downarrow-(\partial\mu/\partial B)_\uparrow$. It is important
that this method of determining the $g$-factor does not require the
use of any fitting procedures or parameters. Figure~\ref{fig8}~(a, b)
shows measured $\partial\mu/\partial B$ as a function of the
deviation of the filling factor from 2 at two values of magnetic
field. Near $\nu=2$, there are sharp intra-level interaction-induced
structures discussed above; these regions have been excluded from the
analysis. However, farther from $\nu=2$, the dependences for $\nu<2$
and $\nu>2$ become parallel to each other. This ensures that the
so-determined $g^*$ is not affected by the valley splitting
\cite{hrapai03,valley} and intra-level interaction effects
\cite{macdonald86,efros88} discussed above. The latter contribute
equally to both spin-up and spin-down dependences and cancel each
other out. Disorder also contributes equally to $\partial\mu/\partial
B$ on both sides of $\nu=2$: we have found that at magnetic fields
down to approximately 3~T, there are wide regions of filling factors
where capacitance ({\it i.e.}, the density of states) is symmetric
around $\nu=2$ (see, {\it e.g.}, Fig.~\ref{fig7}~(a)); furthermore,
closeness of the capacitance to $C_0$ attests that the
disorder-induced corrections are small. At lower magnetic fields,
however, the electron-hole symmetry around $\nu=2$ breaks down, which
sets the lower boundary for the range of magnetic fields (and,
consequently, electron densities). Note that temperature smears out
the dependences in a way similar to disorder: at higher temperatures,
the capacitance at half-integer filling factors decreases, which
leads to a worsening of the method accuracy.

\begin{figure}\centering
\scalebox{0.6}{\includegraphics[clip]{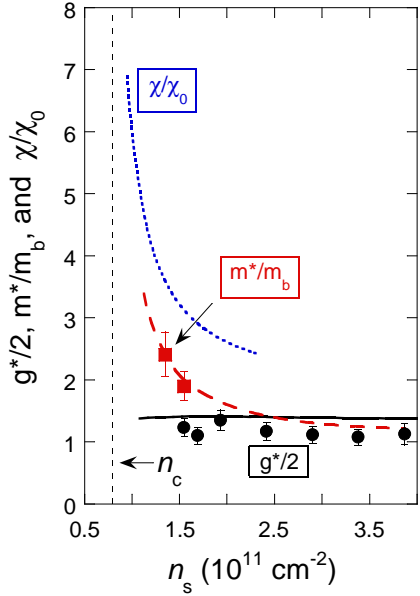}}
\caption{\label{fig9} The effective $g$-factor (circles) and the
cyclotron mass (squares) as a function of the electron density. The
solid and long-dashed lines represent, respectively, the $g$-factor
and effective mass, previously obtained from transport measurements
\cite{shashkin02}, and the dotted line is the Pauli spin
susceptibility obtained by magnetization measurements in parallel
magnetic fields. The critical density $n_c$ for the $B=0$
metal-insulator transition is indicated.}
\end{figure}

In Fig.~\ref{fig9} we plot the measured $g$-factor along with the one
previously obtained from transport measurements (solid line). One can
see that there is no systematic dependence of the $g$-factor on
$n_s$: it remains approximately constant and close to its value in
bulk silicon even at the lowest electron densities, which is in good
agreement with the transport \cite{shashkin02} and magnetocapacitance
\cite{hrapai03} results.

The same method can be used for determination of the cyclotron mass
in tilted magnetic fields strong enough to completely polarize the
electron spins \cite{remark2}. If (and only if) the spin splitting
exceeds the cyclotron splitting, the gap at $\nu=2$ lies between
Landau levels 0$\uparrow$ and 1$\uparrow$, and the difference
$(\partial\mu/\partial B)_{N=1}-(\partial\mu/\partial B)_{N=0}$ is
equal to $2\mu_B\,(m_e/m^*)\cos\,\phi$, where $\phi$ is the tilt
angle. Once the electron spins are fully polarized at filling factors
above $\nu=2$, the tilt angle is automatically large enough for the
level crossing to have occurred. The region of explorable electron
densities is restricted from above by the condition that the
electrons must be fully spin-polarized, while with our current
set-up, the maximum magnetic field at which we can apply the
modulation is only 8~tesla capable of polarizing the electron spins
up to $n_s^*\approx2\times10^{11}$~cm$^{-2}$ (see previous subsection
and Ref.~\cite{vitkalov00}). Figure~\ref{fig8}~(c, d) shows
$\partial\mu/\partial B$ as a function of $|\nu-2|$ under the
condition $n_s<n_s^*$ at two tilt angles \cite{angle}. The extracted
cyclotron mass at electron densities $1.55$ and
$1.35\times10^{11}$~cm$^{-2}$ is significantly enhanced. At densities
below $1.35\times10^{11}$~cm$^{-2}$, the symmetry of capacitance on
both sides of the $\nu=2$ gap breaks down, making the determination
of $m^*$ impossible. As a result, we were only able to obtain two
data points. Nevertheless, good agreement with the effective mass
previously obtained by transport measurements (Fig.~\ref{fig9})
demonstrates the applicability of the new method and adds credibility
to both transport and magnetization results.

We stress once again that the advantage of the new method we use here
is that it allows determination of the spectrum of the 2D electron
system under the condition that the Fermi level lies outside the
spectral gaps, and the inter-level interactions are avoided. Being
symmetric about $\nu=2$, the intra-level interactions are canceled
out in the data analysis and do not influence the extracted
$g$-factor and cyclotron mass. Therefore, the obtained values $g^*$
and $m^*$ are likely to be identical with those for a continuous
spectrum, and the comparison with previously found values of the
$g$-factor and the effective mass is valid.

\section{Conclusions}

The Pauli spin susceptibility has been determined by measurements of
the thermodynamic magnetization and density of states in a
low-disordered, strongly correlated 2D electron system in silicon in
parallel magnetic fields. It is found to behave critically near the
zero-field MIT, which is characteristic of the existence of a phase
transition. Magnetization measurements in perpendicular and tilted
magnetic fields allow determination of the spectrum characteristics
of 2D electron systems and show that enhancement of the $g$-factor is
weak and practically independent of the electron density, while the
cyclotron mass becomes strongly enhanced as the density is decreased.
The obtained data agree well with the $g$-factor and effective mass
obtained by transport measurements, as well as with the Pauli spin
susceptibility obtained by magnetization measurements in parallel
magnetic fields, even though the lowest electron densities reached in
the second experiment are somewhat higher. Thus, we arrive at the
conclusion that, unlike in the Stoner scenario, it is indeed the
effective mass that is responsible for the dramatically enhanced spin
susceptibility at low electron densities. This result is consistent
with conclusions of the recent theory by Punnoose and Finkelstein
\cite{punnoose05} who made a renormalization group analysis for
multi-valley 2D systems and found that the effective mass
dramatically increases at the metal-insulator transition while the
$g$-factor remains nearly intact. We note, however, that the
theoretical spin susceptibility diverges at disorder-dependent
density $n_c$ \cite{punnoose05}, whereas the experimental $\chi$
grows critically near the disorder-independent density $n_\chi$
\cite{review1}.

\section*{Acknowledgments}

We gratefully acknowledge discussions with S. Chakravarty, B.~I.
Halperin, D. Heiman, N.~E. Israeloff, R.~S. Markiewicz, and M.~P.
Sarachik. We would also like to thank A. Gaidarzhy and J.~B. Miller
for technical assistance and C.~M. Marcus and P. Mohanty for an
opportunity to use their nanofabrication facilities. This work was
supported by NSF grant DMR-0403026, PRF grant 41867-AC10, the RFBR,
RAS, and the Programme ``The State Support of Leading Scientific
Schools''.


\begin{thebibliography}{00}
\bibitem{rem} To deal with this problem, the ``subtraction of the
diamagnetic contribution'' was suggested in Ref.~\cite{prus03}. The
diamagnetic contribution $\Delta$ was determined in the
partially-polarized regime as the difference between the direct
magnetization data (``Mag'') and the data obtained from
Shubnikov-de~Haas oscillations (``SdH''):
$\Delta=\mbox{Mag}-\mbox{SdH}$. The experimental data were then
corrected by $\Delta$. We find this procedure meaningless as it
essentially results in replacing the magnetization data by
Shubnikov-de~Haas data:
$\mbox{Mag}-\Delta=\mbox{Mag}-(\mbox{Mag}-\mbox{SdH})=\mbox{SdH}$. In
fact, the difference between magnetization and Shubnikov-de~Haas data
in their experiment is likely to be due to the presence of a band
tail of localized electrons at all electron densities in their
sample.
\bibitem{remark} Since the diamagnetic shift decreases with
increasing $n_s$ and/or decreasing $B$, it may in principle be
noticeable at the lowest $n_s$ and highest $B$ used in the
experiment. Comparing low- and high-field curves shown in
Fig.~\ref{fig3}(a), we estimate that at
$n_s\sim1\times10^{11}$~cm$^{-2}$ and $B=7$~T, the contribution of
the diamagnetic shift is less than $0.2\mu_B$.
\bibitem{rem1} The critical density for the MIT was determined from
transport measurements (see Refs.~\cite{review1,MIT}).
\bibitem{rem2} The fact that $B_c\rightarrow0$ as $n_s\rightarrow
n_c$ speaks in favor of the strongly correlated liquid being close to
the crystal \cite{nozieres79}.
\bibitem{rem3} The geometric capacitance has been determined
according to $1/C_0=1/C|_{B=0}-1/Ae^2\left.(dn_s/d\mu)\right|_{B=0}$
(for more on this procedure, see Ref.~\cite{hrapai03}).
\bibitem{valley} The bare valley splitting is independent of the
magnetic field and does not contribute to $\partial\mu/\partial B$.
However, it may be enhanced by inter-level interactions
\cite{hrapai03}. The fact that $\partial\mu/\partial B$ for $\nu<2$
and $\nu>2$ are parallel to each other ensures that a possible
influence of the enhanced valley splitting is negligible in our
experiment.
\bibitem{remark2} The effects of finite layer thickness, which lead to
an increase of the effective mass with parallel magnetic field, are
negligible in silicon MOSFETs \cite{review1}.
\bibitem{angle} Using $n_s^*(8~\mbox{T})\approx2\times10^{11}$~cm$^{-2}$, one can estimate that the coincidence of spin and cyclotron splittings for $\nu=2$ occurs at $\phi^*=\cos^{-1}(n_s^*hc/2eB)\approx59^\circ$.
\end{thebibliography}
\end{document}